\documentclass[aps,prb,twocolumn,superscriptaddress,showpacs]{revtex4-1}%superscriptaddress
\usepackage{graphicx}
\usepackage{amsmath,amssymb,bm}
\usepackage[colorlinks,citecolor=blue,linkcolor=blue]{hyperref}

\begin{document}

\title{Anomalous Energy Transport across Topological Insulator Superconductor Junctions}

\author{Jie Ren}\email{renjie@lanl.gov}
\affiliation{Theoretical Division, Los Alamos National Laboratory, Los Alamos, New Mexico 87545, USA}

\author{Jian-Xin Zhu}\email{jxzhu@lanl.gov}
\affiliation{Theoretical Division, Los Alamos National Laboratory, Los Alamos, New Mexico 87545, USA}
\affiliation{Center for Integrated Nanotechnologies, Los Alamos National Laboratory, Los Alamos, New Mexico 87545, USA}

\date{\today}

\begin{abstract}
We study the nonequilibrium energy transport across a topological insulator/superconductor junction, by deriving an interfacial heat current formula through scattering wave approach. Several anomalous thermal properties are uncovered, such as thermal energy's Klein tunneling, asymmetric Kapitza resistance and negative differential thermal resistance. We expect these findings could have potential applications for the energy control in various hybridized mesoscopic systems.
\end{abstract}

% 73.40.-c Electronic transport in interface structures 
% 68.35.Ja Surface and interface dynamics and vibrations  
% 73.20.-r Electron states at surfaces and interfaces  
% 73.50.Bk General theory, scattering mechanisms  
% 74.78.-w Superconducting films and low-dimensional structures  
% 05.60.Gg Quantum transport  

% 66.70.-f Nonelectronic thermal conduction and heat-pulse propagation in solids; thermal waves (for thermal conduction in metals and alloys, see 72.15.Cz and 72.15.Eb)
% 72.10.Bg, General formulation of transport theory
% 03.65.Vf Phases: geometric; dynamic or topological
% 44.10.+i Heat conduction (see also 66.60.+a and 66.70.-f in transport properties of condensed matter)
\maketitle

Topological insulators (TIs), characterized by a bulk gap and a gapless surface mode with a Dirac-like linear dispersion, are presently one of the most interesting topics in condensed matter physics~\cite{Hasan2010RMP, Qi2011RMP}. Their conducting surface states in the insulating gap are topologically protected by time-reversal symmetry, hence robust to disorders and perturbations, potentially leading to various device applications. With the help of doping-induced superconductivity in TIs~\cite{Hor2010PRL} or by depositing superconducting materials on TIs due to the proximity effect~\cite{Fu2008PRL, Fu2009PRL, Qi2010PRB},  the interplay between the superconducting ordering and the gapless chiral surface state has triggered much interest~\cite{Qi2011RMP}.  

On the one hand, such topological insulator superconductor (TI/S) junctions have been used to create chiral Majorana fermions for topological quantum computations and for the study of their impacts on electronic tunneling properties \cite{Akhmerov2009PRL, Tanaka2009PRL,  Law2009PRL, Linder2010PRL, Ioselevich2011PRL}.  However, the thermal properties of such systems have not yet paid equal attention. 
On the other hand, hybrid superconducting circuits are widely used for quantum computing and simulations \cite{you1, you2}, where managing heat dissipation at cryogenic temperatures becomes important for the right device operation.
%We turn attention to the thermal energy transport in TI/S junctions.
Thus, in view of the fact that hybrid topological superconductor junctions could be a natural candidate for quantum simulations and computing, the bottleneck in the future will be efficiently managing heat dissipation/refrigeration and controlling energy transport in such systems. Therefore,  understanding thermal properties of such hybrid mesoscopic structures at cryogenic temperatures will be crucial for future quantum/nano technology~\cite{Giazotto2006RMP, super_heat_review}. It even has the potential to open up a rich variety of thermal device concepts in superconducting-circuit-based hybrid systems, just like thermal diodes, thermal transistors, thermal logic gates and thermal memories in dielectric phononics~\cite{Li2012RMP}. 

In this paper, we study the nonequilibrium energy transport across a TI/S junction interface and uncover its anomalous thermal properties, such as thermal energy's Klein tunneling, asymmetric Kapitza resistance and negative differential thermal conductance (NDTC).
Among them,  Kapitza resistance measures the interfacial thermal resistance when thermal energy flows through the interface between two different materials \cite{Pollack1969RMP, Swartz1989RMP}. Asymmetric Kapitza resistance is one unusual thermal property that the interface acts as a good thermal conductor if a positive thermal bias is applied, while with a negative thermal bias it exhibits poor thermal conduction, thus effectively acting as a thermal insulator. As such,  it functions as a thermal rectifier or diode  (for a review, see \cite{Li2012RMP, Roberts2011IJTS}).
While the NDTC, another unusual thermal transport phenomenon, where the heat current across a thermal conductor decreases when the temperature bias increases (for a review, see \cite{Li2012RMP}), is an essential element for the construction of thermal transistors and thermal logic gates, and has been shown to exist in many anharmonic lattice systems. These concepts were usually restricted to the pure phononic systems, where the thermal energy is carried by quantized lattice vibrations, phonons. Here, we report similar findings in superconducting hybridized mesoscopic junctions, which could extend the conceptual thermal devices in dielectrics into superconductor-based hybrid systems and could have great potential applications at cryogenic temperatures \cite{cooling} in the near future.

\begin{figure}%[t]
\vspace{-1.2cm}
%\scalebox{0.31}[0.31]{\includegraphics{scheme.eps}}
\scalebox{0.35}[0.32]{\includegraphics{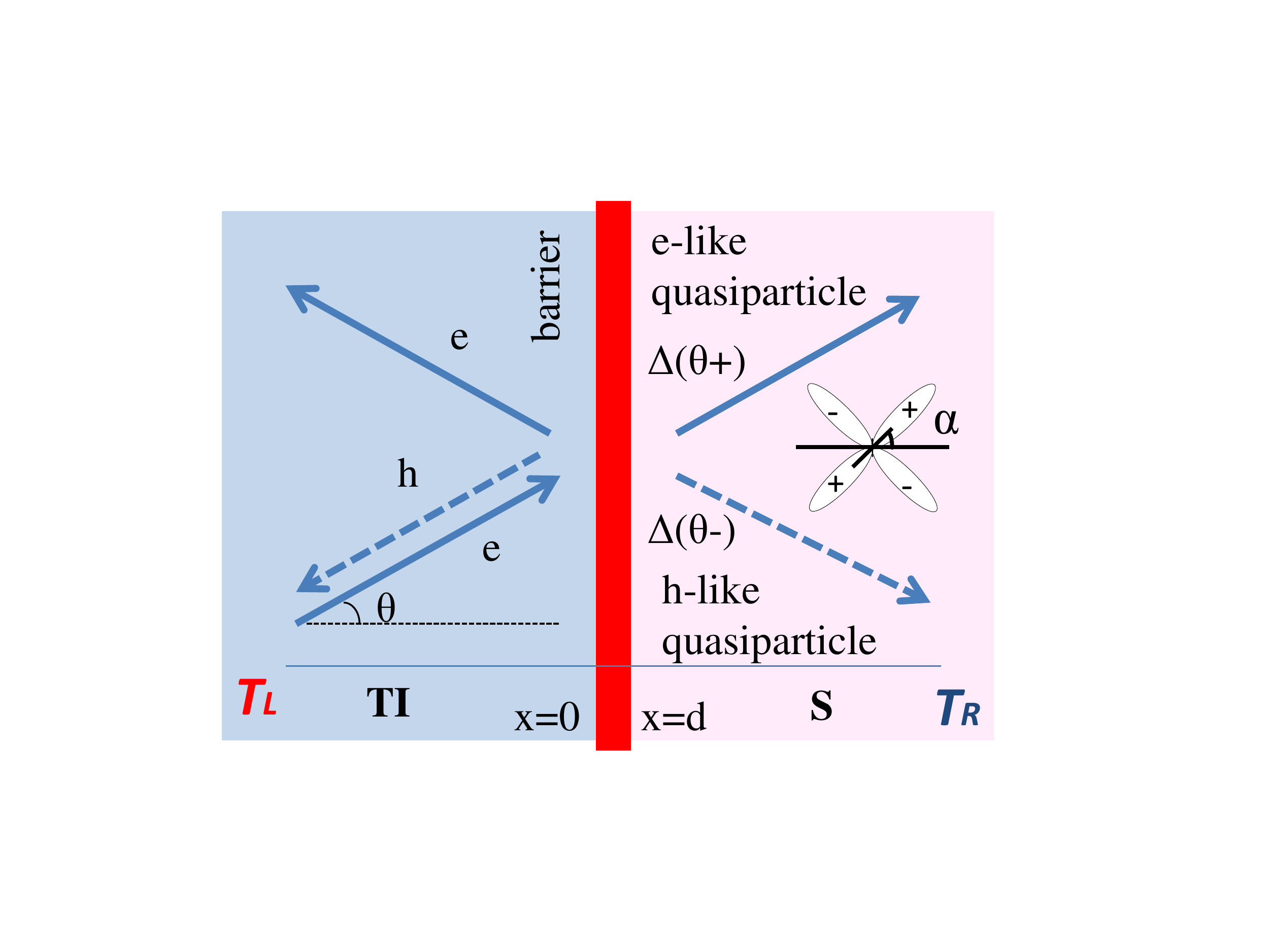}}
\vspace{-1.8cm}
\caption{(color online). Schematic illustration of the reflections and transmissions at the TI/S interfaces.}
\label{fig:scheme}
\end{figure}

As depicted in Fig. \ref{fig:scheme}, we consider a two dimensional (2D) TI/S junction attached to their respective thermal reservoirs at $T_L$ and $T_R$, with an insulating barrier locating at $x\in[0,d]$. The 2D TI junction part ($x<0$) could be formed on the surface of 3D topological insulators, the S junction part ($x>d$) could be induced via the proximity effect, and the barrier (B) part could be constructed by gating voltage or doping. The bulk of TI is a bad insulator of electrons and when doped or with other disorders, both electron and phonon  will be strongly scattered but with the surface electronic state unaffected (so called topologically protected). Thus, the bulk contribution to the energy transport will be seriously suppressed and negligible. Moreover, the phonon contribution is not only suppressed seriously by the doping and disorders in the bulk, but also blocked by the barrier layer located at the interface between TI and S sides. When TI and S sides have large lattice mismatch, the phonon contribution will be further reduced. Therefore, in the present work we only focus on the thermal transport contributed by the surface metallic states.

We assume the same Fermi level in both TI and S, and employ the Bogoliubov-de Gennes (BdG) equation $\hat H\psi=E\psi$ to study the thermal transport properties. The Hamiltonian of the surface state on the topological superconductor is given by \cite{Hasan2010RMP, Qi2011RMP}
\begin{eqnarray}
&&\hat{H}=\left(
        \begin{array}{cc}
          \hat{H}_0(\textbf{k})-E_{F} & \hat\Delta \\
          -\hat\Delta^{*} & -\hat{H}^*_0(-\textbf{k})+E_{F} \\
        \end{array}
      \right)\label{Hamiltonian}
\end{eqnarray}
acting on a Nambu basis $(\psi_{e\uparrow}, \psi_{e\downarrow}, \psi_{h\uparrow}, \psi_{h\downarrow})$, with $\hat{H}_0(\textbf{k})=\hbar v_{F}(k_x \hat\sigma_y-k_y\hat\sigma_x)+U_0\Theta(x)\Theta(d-x)$ and $\hat\Delta=i\hat\sigma_y\Delta(\theta,T)\Theta(x-d)$. Here, $v_F$ is the Fermi velocity, $\hat\sigma_{x(y)}$ denote Pauli matrices, $U_0$ is the barrier potential and $\Delta(\theta,T)$ depicts the order parameter with a given pairing symmetry and temperature dependencies. Throughout the work, we consider $E_F\gg (|\Delta|, E)$ to satisfy the mean-field nature of BdG approach, {\it i.e.}, the superconducting coherence length $\hbar v_F/|\Delta|$ is much larger compared to the Fermi wavelength $\hbar v_F/E_F$.
%a weak coupling approximation $E_F\gg |\Delta|$. 
%$E$ to satisfy the mean-field nature of BdG approach, which is valid only when the superconducting coherence length $\hbar v_F/|\Delta|$ is much large compared to the Fermi wavelength $\hbar v_F/E_F$. 
Adjusting $E_F$ can be achieved via doping or gate voltage.

Since in superconductor junctions, the quasiparticles are carriers of thermal energy, we need to obtain the transmission (equivalently the reflection) coefficients of quasiparticles in order to investigate the thermal transport properties. Considering the conservation law for particle current, we can simplify the problem by merely considering the particle (including both electron and hole) current in the side of TI. Defining $f=\binom{\psi_{e\uparrow}}{\psi_{e\downarrow}}$ and its hole counterpart $g=\binom{\psi_{h\uparrow}}{\psi_{h\downarrow}}$, we have the probability density for finding either an electron or a hole, $P=|f|^2+|g|^2$. By using the BdG equation $i\hbar\partial_t\binom{f}{g}=\hat{H}\binom{f}{g}$ with Eq. (\ref{Hamiltonian}), and considering the continuity equation $\partial_t P+\nabla \cdot J_P=0$, we obtain the $x$-component of the particle current:
\begin{equation}\label{DiracJ}
J^x_P=v_F(f^{\dag}\hat\sigma_yf-g^{\dag}\hat\sigma_yg)\;.	
\end{equation}
Note the hole current is a time-reversed counterpart of the electron contribution so that it naturally obtains an opposite sign compared with the electron current. 
If we express the whole wave function at the TI side, $\Psi_{\text{TI}}:=\binom{f}{g}$, in terms of the normal reflection amplitude, $b$, and Andreev reflection~\cite{BTK1} (for electron-hole conversion)  amplitude $a$, which will be defined explicitly below, and substitute $\binom{f}{g}$ into Eq. (\ref{DiracJ}), we then obtain the particle current as
\begin{equation}
J^x_P(E,\theta)=2v_F\cos\theta\left(1-|a(E,\theta)|^2-|b(E,\theta)|^2\right).
\end{equation}
This expression of the particle current has a clear physics picture that $\tilde\kappa(E,\theta):=1-|a(E,\theta)|^2-|b(E,\theta)|^2$ denotes the transmission of energy carriers with energy $E$ and incident angle $\theta$, $v_F\cos\theta$ the effective velocity in the $x$ direction and $2$ the spin degeneracy. Considering the carrier's energy $E$, the Fermi occupation difference between two sides of the interface $f_L-f_R=\frac{1}{e^{{E}/{k_BT_L}}+1}-\frac{1}{e^{{E}/{k_BT_R}}+1}$ and summation over all possible incidence angles and momenta $\sum_{k}\int^{\pi/2}_{-\pi/2}d\theta=\frac{1}{2\pi}\int dE\int^{\pi/2}_{-\pi/2}d\theta (dE/dk)^{-1}=\int dE\int^{\pi/2}_{-\pi/2}d\theta \frac{1}{2\pi\hbar v_F}$, we arrive at the energy current expression:
\begin{eqnarray}\label{JQ}
J_Q&=&%\iint dEd\theta\frac{J^x_P}{hv_F}=
\frac{2}{h}\int^{\infty}_{-\infty}dE E\kappa(E)[f_L-f_R]\;, 
\end{eqnarray}
with $\kappa(E)=\int^{\pi/2}_{-\pi/2}d\theta \cos\theta \tilde\kappa(E,\theta)=\int^{\pi/2}_{-\pi/2}d\theta \cos\theta (1-|a|^2-|b|^2)$.
Note that this expression is general and is obtained before solving the reflection coefficients $a, b$. In fact, similar expressions of the energy current in 1D topological-trivial metal superconductor junctions have been obtained by a rigorous derivation through linear response of entropy production~\cite{Riedel1993PRB} or by a heuristic argument \cite{Bardas1995PRB}. The latter was then applied to the superconducting graphene systems~\cite{graphene1,graphene2}.  It is worth emphasizing that all the transports considered in this work are charge neutral, \emph{i.e.}, the carriers transport only thermal energy without charge current. In fact, if we follow the same procedure for charge transport, we will arrive at a similar expression for electric current: $J_e=\frac{2e}{h}\int dE \int d\theta \cos\theta \tilde\sigma(E,\theta)[f_L-f_R]$, with $\tilde\sigma=1+|a|^2-|b|^2$, similar to the BTK formula~\cite{BTK1, BTK2}.  One can then get that due to the even symmetry of $\tilde\sigma$ ($\tilde\sigma(E,\theta)=\tilde\sigma(-E,\theta)$) and the odd symmetry of $f_L-f_R$ with respect to $E$, $J_e$ vanishes as zero. This is also a consequence of the particle-hole symmetry in our system.

We now proceed to determine the scattering coefficients of Andreev reflection amplitude $a$ and normal reflection amplitude $b$ by imposing the boundary conditions on the  wave functions   at the interfaces of barrier. Diagonalizing Eq. (\ref{Hamiltonian}) straightforwardly yields the wave functions in the TI, barrier, S regions. In the TI region $(x<0)$, for electrons and holes traveling the $\pm x$ direction with a conserved transverse momentum $k_y$ and energy $E$ measured from $E_F$, the wave functions are given as
\begin{eqnarray*}
\psi^{e\pm}_{\text{TI}}&=&(1,\pm ie^{\pm i\theta},0,0)e^{i(\pm k_xx+k_yy)},  \\ \psi^{h\pm}_{\text{TI}}&=&(0,0,i,\pm e^{\pm i\theta})e^{i(\mp k_xx+k_yy)},
\end{eqnarray*}
where $k_x=E_F \cos\theta/(\hbar v_F)$ and $\theta=\arcsin(\hbar v_F k_y/E_F)$ is the angle of incidence. Note we have used the mean-field condition $E_F\gg E$. In the barrier region $(0<x<d)$, we employ the thin barrier limit \cite{Bhattacharjee2006PRL}: $d\rightarrow0$, $U_0\rightarrow\infty$ but with a constant product $Z\equiv U_0d/(\hbar v_F)$, characterizing the strength of the insulating barrier,  we then obtain
\begin{eqnarray*}
\psi^{e\pm}_{\text{B}}&=&(1,\pm i,0,0)e^{i (\pm Zx/d+k_yy)}, \\
\psi^{h\pm}_{\text{B}}&=&(0,0,i,\pm 1)e^{i (\mp Zx/d+k_yy)}.
\end{eqnarray*}
In the S region $(x>d)$, the electron-(hole-)like quasiparticles are mixtures of electrons and holes. Thus, the transmitted wave functions have the forms:
\begin{eqnarray*}
\psi^{e+}_{\text{S}}&=&(1,ie^{i\theta},-i\Gamma_+e^{i(\theta-\phi_+)},\Gamma_+e^{-i\phi_+})e^{i(k_xx+k_yy)}, \\ \psi^{h+}_{\text{S}}&=&(\Gamma_-,-i\Gamma_-e^{-i\theta},ie^{-i(\theta+\phi_-)},e^{-i\phi_-})e^{i(-k_xx+k_yy)}, \end{eqnarray*}
where $e^{i\phi_{\pm}}={\Delta(\theta_{\pm},T)}/{|\Delta(\theta_{\pm},T)|}$ with $\theta_+=\theta, \theta_-=\pi-\theta$ and $\Gamma_{\pm}=v_{\pm}/u_{\pm}$, with $u^2_{\pm}=\frac{1}{2}(1+\sqrt{E^2-|\Delta(\theta_{\pm},T)|^2}/|E|)=1-v^2_{\pm}$. %For isotropic $s$ wave, $\Delta(\theta,T)=\Delta(T)$ while
For a  $d$-wave pairing symmetry, $\Delta(\theta,T)=\Delta(T)\cos(2\theta-2\alpha)$ with $\Delta(T)=\Delta_0\tanh[(\pi k_BT_c/\Delta_0)\sqrt{0.953(T_c/T-1)}]$ \cite{Tao2012PRB}. Here $T_c$ is the critical temperature, $\Delta_0$ denotes the superconducting gap at zero temperature and $\alpha$ is the angle between the normal direction of the barrier interface and the $x$ axis of the $d_{x^2-y^2}$-wave superconductor.
By taking into account the boundary conditions:
\begin{equation*}
\Psi_{\text{TI}}|_{x=0}=\Psi_{\text{B}}|_{x=0}, \quad\quad \Psi_{\text{B}}|_{x=d}=\Psi_{\text{S}}|_{x=d},
\end{equation*}
with $\Psi_{\text{TI}}=\psi^{e+}_{\text{TI}}+b\psi^{e-}_{\text{TI}}+a\psi^{h-}_{\text{TI}}$, $\Psi_{\text{S}}=t_e\psi^{e+}_{\text{S}}+t_h\psi^{h+}_{\text{S}}$,
$\Psi_{\text{B}}=r_1\psi^{e+}_{\text{B}}+r_2\psi^{e-}_{\text{B}}+r_3\psi^{h+}_{\text{B}}+r_4\psi^{h-}_{\text{B}}$,
the Andreev and normal reflection coefficients are found to be
%\begin{widetext}
\begin{eqnarray*}
a&=&\frac{-\cos^2\theta \Gamma_+ e^{i(\theta-\phi_+)}}{\cos^2\theta+\sin^2Z\sin^2\theta(1-\Gamma_+\Gamma_-e^{i(\phi_--\phi_+)})}, \\
b&=&\frac{\sin Z\sin\theta(\cos Z\cos\theta-i\sin Z)(1-\Gamma_+\Gamma_-e^{i(\phi_--\phi_+)})}{-e^{-i\theta}[\cos^2\theta+\sin^2Z\sin^2\theta(1-\Gamma_+\Gamma_-e^{i(\phi_--\phi_+)})]}.
\end{eqnarray*}
%\end{widetext}
Finally, using the obtained coefficients $a$ and $b$, we get
\begin{eqnarray}
\tilde{\kappa}(E,\theta):&=&1-|a(E,\theta)|^2-|b(E,\theta)|^2 \nonumber\\
&=&\tilde{\kappa}_{\text{TI}}
\frac{1-\tilde{\kappa}_{\text{TI}}|\Gamma_+|^2+(\tilde{\kappa}_{\text{TI}}-1)|\Gamma_+\Gamma_-|^2}
{|1+(\tilde{\kappa}_{\text{TI}}-1)\Gamma_+\Gamma_-e^{i(\phi_--\phi_+)}|^2},
\label{eq:kkk}
\end{eqnarray}
where
\begin{eqnarray}
\tilde\kappa_{\text{TI}}:=\tilde{\kappa}(|E|\gg\Delta_0,\theta)=
\frac{\cos^2\theta}{\cos^2\theta+\sin^2Z\sin^2\theta}
\label{eq:Klein}
\end{eqnarray}
is reminiscent of the relativistic Klein tunneling~\cite{Klein}, as a consequence of  the spin-orbit coupling  in TIs. The barrier becomes transparent  for the thermal energy transport at the resonance condition, $Z:=\frac{U_0d}{\hbar v_F}=n\pi, n=0,\pm1,\cdots$ (such that $\sin Z=0$), or at the normal incidence ($\theta=0$). Eqs. (\ref{eq:kkk}) and (\ref{eq:Klein}) are one of the main results, which enable us to uncover in the following the anomalous thermal properties, such as thermal energy's Klein tunneling, asymmetric Kapitza resistance and NDTC.

\begin{figure}%[t]
\hspace{-.4cm}
%\scalebox{0.31}[0.31]{\includegraphics{scheme.eps}}
\scalebox{0.42}[0.42]{\includegraphics{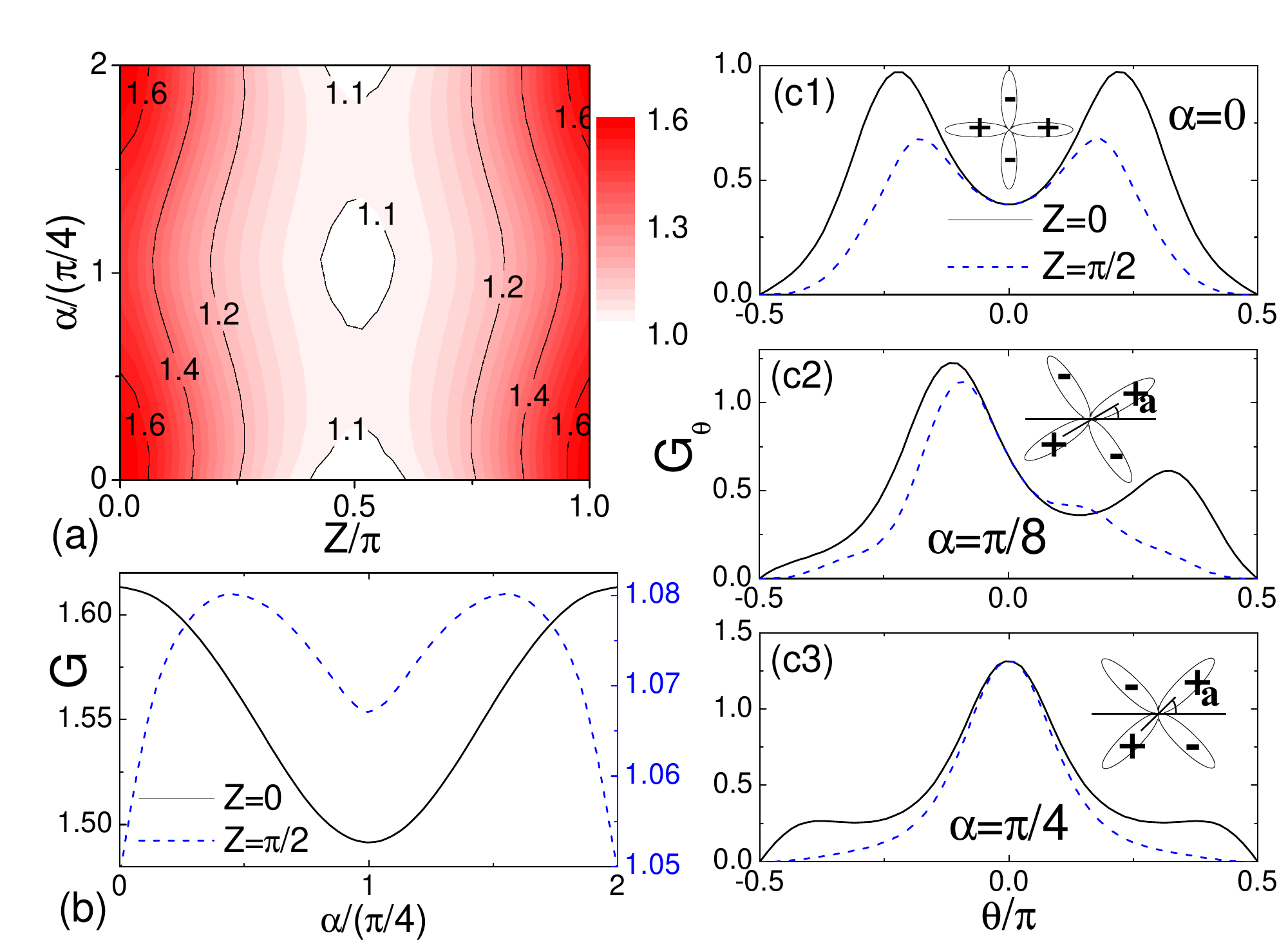}}
\vspace{-.4cm}
\caption{(color online). Thermal energy's Klein tunneling. (a) Thermal conductance as a function of the barrier strength $Z$ and the rotation angle $\alpha$ of the $d$-wave order parameter. (b)  Intersections of (a) for $Z=0$, and $\pi/2$. (c) Angle-resolved thermal conductance $G_{\theta}$ for different $\alpha$ and $Z$. Solid lines are for $Z=0$ while dashed ones for $Z=\pi/2$. Parameters are $T=70$K, $T_c=116$K, $\Delta_0=30$meV. Thermal conductance $G$ is in the unit of ${\Delta_0k_B}/{h}$.}
\label{fig:G}
\end{figure}

In the linear response regime $T_L=T+\delta T/2, T_R=T-\delta T/2, \delta T\rightarrow0$, we have the thermal conductance:
\begin{equation}
G:=\frac{J_Q}{\delta T}=\frac{2}{h}\int^{\infty}_{-\infty}dE \int^{\pi/2}_{-\pi/2}d\theta \frac{E^2 \cos\theta \tilde\kappa(E,\theta)}{4k_BT^2\cosh^2(\frac{E}{2k_BT})}\;.
\label{eq:G}
\end{equation}
As illustrated in Fig.~\ref{fig:G}(a), the oscillatory dependence of $G$ on the barrier strength $Z$ shows the Klein tunneling behavior of energy: thermal conductance anomalously increases when the barrier increases within $[(n+1/2)\pi, (n+1)\pi]$, consistent with the behavior of $\sin^{2}Z$ in Eq. (\ref{eq:Klein}). $G$ also has an oscillatory dependence on $\alpha$, the rotation angle of the superconducting order parameter. As detailed in Fig.~\ref{fig:G}(b), when $Z=0$, $G$ monotonically decreases as $\alpha$ rotates from $0$ to $\pi/4$, but when $Z$ increases to $\pi/2$, $G$ first increases and then decreases. This nonmonotonic behavior is due to the competition of Klein tunneling and superconducting order parameter (gap), which we explain below.

In the S part, thermal energy is carried by the quasiparticles, which only  transport beyond the gap $\Delta$, so that the smaller $\Delta$, the larger thermal conductance. In other words, the rotation angle $\alpha$ adjusts $G$ through adjusting the angle dependence of the superconducting gap. To further understand the competition between Klein tunneling and order parameter $\Delta$, we define the angle-resolved thermal conductance $G_{\theta}$ as $G=\int^{\pi/2}_{-\pi/2} d\theta G_{\theta}$ and plot it as a function of the incidence angle $\theta$ in Figs. \ref{fig:G}(c1-c3). For the case of $Z=0$, when $\alpha=0$, $G_{\theta}$ has two peaks around $\theta=\pm{\pi}/{4}$ where $\Delta$ is gapless; while, when $\alpha$ rotates to $\pi/4$, two peaks becomes a single peak around $\theta=0$. As a result, $G$, the angle integration of $G_{\theta}$, decreases as $\alpha$ increases  from $0$ to $\pi/4$. For $\alpha$ changing from  $\pi/4$ to $\pi/2$, the behavior is symmetrically reversed.  When the barrier $Z$ increases, the Klein tunneling comes into play. For the case of $Z=\pi/2$, when $\alpha=0$, although there are still two peaks for thermal conductance, their intensity is suppressed dramatically. When $\alpha=\pi/4$, the gapless angle coincides with the normal incidence angle, the barrier becomes transparent. Therefore, the Klein tunneling helps to keep the single conductance peak (at $\theta=0$) intensity unchanged. At the intermediate regime $\alpha=\pi/8$,  one peak near $\theta=0$ preserves while the other peak far from $\theta=0$ is repressed. As a consequence, the thermal conductance $G$ increases first and then decreases within $\alpha\in[0,\pi/4]$ [see Fig.~\ref{fig:G}(b)]. It is also interesting to notice that at certain angles, the angle-resolved thermal conductance $G_{\theta}$ can be even enhanced by the nonzero barrier strength [see Fig.~\ref{fig:G}(c2)]. This anomalous behavior is a consequence of the competition of thermal Klein tunneling and the orientation angle of $d$-wave superconductor.
We note these predictions would be validated by the present techniques
of angle resolved thermal transport measurements (for a review, see Ref.~\cite{angle}).

\begin{figure}[htp]
%\hspace{-.4cm}
\vspace{.2cm}
\scalebox{0.3}[0.3]{\includegraphics{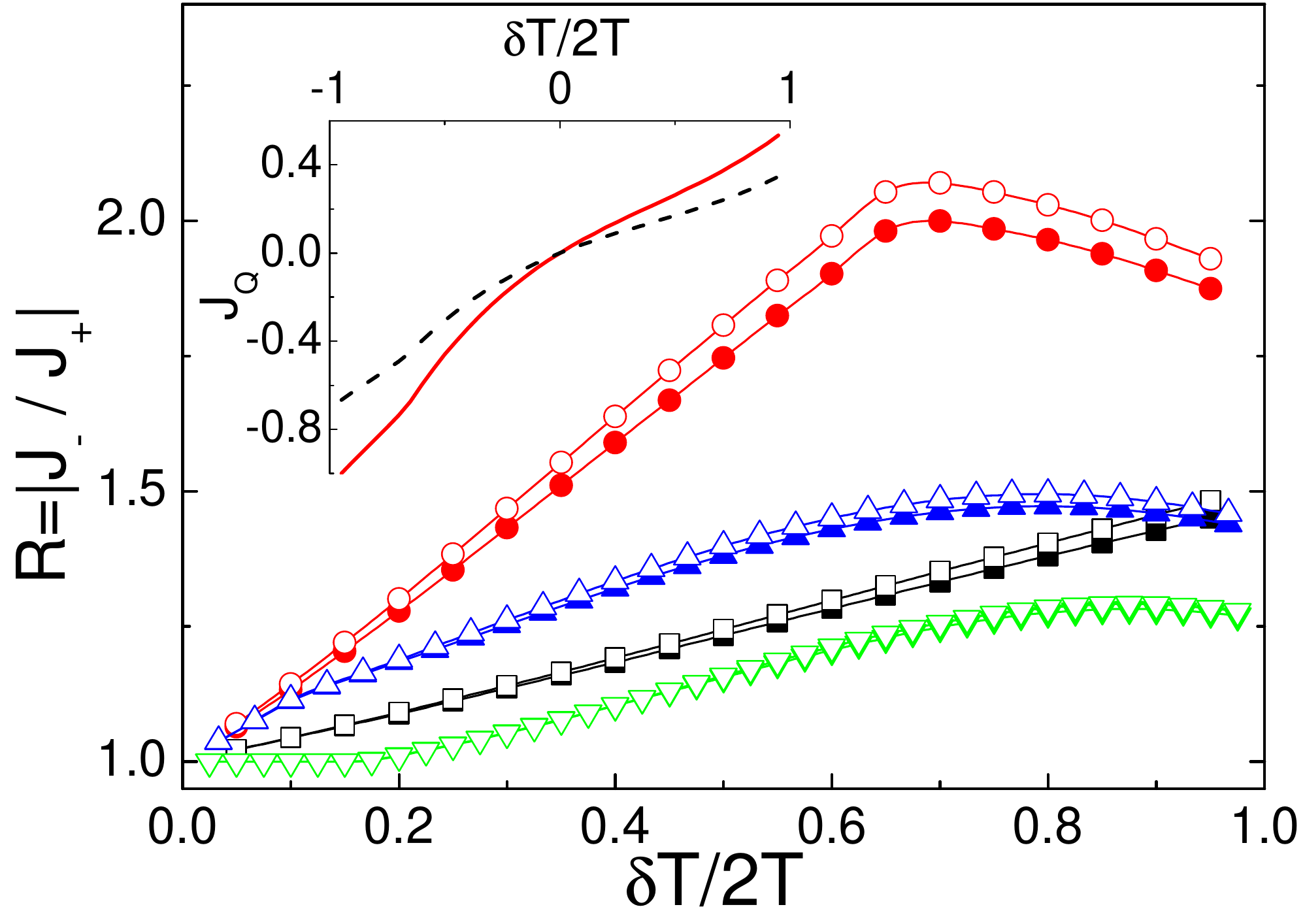}}
\vspace{-.4cm}
\caption{(color online). Rectification ratios as a function of temperatures. Values apart from $1$ indicate asymmetric Kapitza resistances. $\blacksquare$ ($T=35$K), ${\color{red}\bullet}$ ($T=70$K), ${\color{blue}\blacktriangle}$ ($T=104$K), ${\color{green}\blacktriangledown}$ ($T=140$K) denotes for $Z=0$. Their hollow counterparts are for cases of $Z=\pi/2$. Inset shows one example of the $J_Q$ profiles depending on temperature bias for $Z=0$ (solid) and $\pi/2$ (dashed), at $T=70$K. $\alpha=0$. Other parameters are the same as in Fig. \ref{fig:G}. The energy current $J_Q$ is in the unit of $\Delta_0^2/h$.}
\label{fig:diode}
\end{figure}

The asymmetric Kapitza resistance is essentially a nonlinear response behavior.  It is a consequence of different temperature responses of different materials at two sides of the interface \cite{Li2012RMP}. The inset of Fig.~\ref{fig:diode} shows typical $J_Q$ behaviors at $T=70$K for $Z=0$ (solid line) and $\pi/2$ (dashed line), via varying temperature bias. As a measure of the asymmetry, we define the rectification ratio $R= | J_{-}/J_{+} |$, where $J_{+}$ refers to thermal current when $\delta T=T_L-T_R>0$ while $J_{-}$ refers to thermal current after switching the temperature bias $T_L\leftrightarrow T_R$. As shown in Fig.~\ref{fig:diode}, when temperature bias is apart from $0$, $R$ becomes more deviating from 1. Except for the low temperature case, (e.g.  $T=35$K), other three examples show that increasing bias does not always increasing the rectification ratio at the large bias regime. The results also indicate that although the insulating barrier changes the $J_Q$ profiles quite noticeably, it does not change the rectification ratio significantly, which is even slightly enhanced by the barrier. In addition, $R$ has a nonmonotonic temperature dependence that increasing $T$ first increases and then decreases $R$, as exemplified by the highest curves for $T=70$K in Fig. \ref{fig:diode}.
This is reasonable that the asymmetric Kapitza resistance in our system results from the different temperature responses of topological insulator and superconductor at the sides of the interface: At higher temperature, the superconducting gap diminishes so that the superconductor tends to topological insulator and both sides of the interface tend to have the same temperature response, which explains the reduction in $R$. At lower temperatures, although two sides of the interface have distinct temperature responses, the bias $\delta T$ cannot be larger than $T$ such that small bias reduces $R$. Therefore, the optimal $R$ appears at intermediate temperature.

\begin{figure}[htp]
%\hspace{-.4cm}
\vspace{.2cm}
\scalebox{0.3}[0.3]{\includegraphics{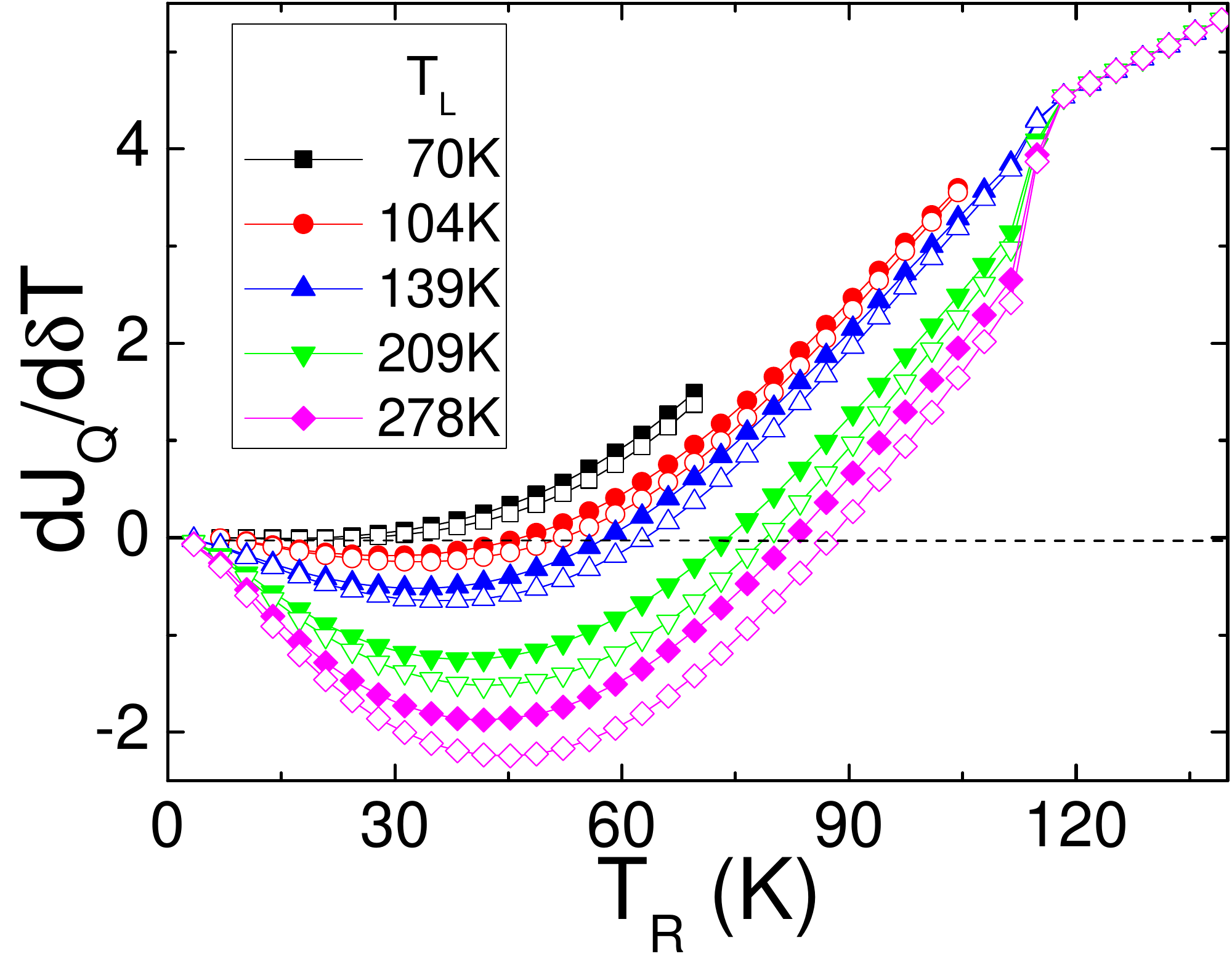}}
\vspace{-.4cm}
\caption{(color online). NDTC for various temperatures at $\alpha=0$ (filled symbols) and  $\alpha=\pi/4$ (hollow symbols). $Z=0$. Other parameters are the same as in Fig.~\ref{fig:G}. The DTC $dJ_Q/d\delta T$ is in the unit of $\Delta_0k_B/h$.}
\label{fig:NDTC}
\end{figure}

The superconductor is usually a bad thermal conductor since its gap $\Delta$ forbids the existence of quasiparticles that is responsible for the energy transport. Increasing the temperature $T_R$ at the S part could increase the energy transmission $\kappa(E)$ (in Eq.~\ref{JQ}) by diminishing the superconducting gap to allow more thermal energy carriers, while simultaneously decrease the temperature bias $\delta T=T_L-T_R$ as well as the occupation difference $f_L-f_R=\frac{1}{e^{{E}/{k_BT_L}}+1}-\frac{1}{e^{{E}/{k_BT_R}}+1}$.  Once the increased $\kappa(E)$ is able to compensate the loss in $f_L-f_R$ due to the decreased $\delta T$, we expect to observe NDTC, {\emph i.e.}, the energy current increases as the bias $\delta T$ decreases. To characterize this anomalous behavior, we define the differential thermal conductance (DTC): $dJ_Q/d\delta T$, for given finite $\delta T$. Note that this is a nonlinear quantity similar to the nonlinear differential electrical conductance $dI/dV$, and is different from the previous defined thermal conductance $G$ in Eq.~(\ref{eq:G}), which is a linear quantity at $\delta T\rightarrow 0$. 

As a showcase, we calculate DTC in Fig.~\ref{fig:NDTC}, with varying $T_R$ in the S region but fixing $T_L$ as a reference temperature. Indeed, NDTC appears as we expect, although it is absent for the low $T_L$ case  ({\it e.g.} $T_L=70K>T_R$). Increasing $T_L$ enhances the regime of NDTC. In addition, tuning the angle of superconducting paring symmetry $\alpha$ can also slightly enhance NDTC, as shown in Fig.~\ref{fig:NDTC}. However, when $T_R$ increases across a threshold and approaches to $T_L$,  NDTC disappears. 
When the superconducting part is replaced with a topological insulator, there exists no NDTC. In fact, when $T_R>T_c$ (see 116 K in Fig.~\ref{fig:NDTC}), the S part fades into non-superconducting TI, which causes the collapse of DTC beyond $T_c$. In this case, we can have  a constant thermal transmission $\kappa(E)\approx\kappa$, where $\kappa$ depends on the barrier strength. 
Then from Eq.~(\ref{JQ}), we have DTC: $dJ_Q/d\delta T=-dJ_Q/dT_R=\frac{2\kappa}{h}\int dE \frac{E^2}{4k_BT^2_R\cosh^2[E/(2k_BT_R)]}=\frac{2\pi^2k^2_B}{3h}\kappa T_R$, which explains the linear behavior beyond the critical temperature in Fig.~\ref{fig:NDTC}.

Finally, we would like to point out that although we exemplified the anomalous thermal transport by a $d$-wave superconductor, in principle, when replaced with a $s$-wave superconductor, the results will be qualitatively the same. The asymmetric Kapitza and NDTC are consequences of the different temperature responses of both the nonsuperconducting TI side and the superconducting side. The thermal energy's Klein tunneling restuls from the Dirac-like linear dispersion of the materials. When competing with the orientation angle of the $d$-wave symmetry, the behavior of angle-resolved thermal transport becomes rich.

In summary, using the scattering wave approach, we have derived an interfacial heat current formula in a TI/S junction. With the help of this formula, we have studied the nonequilibrium energy transport across this interfacial system and have uncovered several anomalous thermal properties for the TI/S interface, such as thermal energy's Klein tunneling, asymmetric Kapitza resistance and negative differential thermal resistance. 

The asymmetric Kapitza resistance and NDTC are already discussed in dielectric phonon systems at high (room) temperatures \cite{Li2012RMP}. But previous studies about the asymmetric resistance and NDTC focus on pure phononic systems, and do not involve any superconductors as well as topological insulators. Here, we have uncovered the anomalous thermal transport in hybrid topological insulator superconductor system. 

One immediate advantage of this kind of hybrid system is that the thermal energy's Klein tunneling renders highly efficient heat dissipation even when the TI/S interface is not perfect that produces a large barrier. However, in a normal metal superconductor system, the heat dissipating ability will be severely reduced by the large barrier induced by the imperfect interface.

Since hybrid superconductor/topological insulator systems are crucial for future quantum/nano technology at cryogenic temperatures,
we believe that understanding the anomalous heat transport in such hybrid  systems would be useful for managing heat dissipation in future cryogenic devices and even reveal the potential applications of such hybrid systems for the smart energy control at mesoscopic scales.

This work was supported by the National Nuclear 
Security Administration of the U.S. DOE  at  LANL under Contract 
No. DE-AC52-06NA25396,  the LDRD Program at LANL (J.R.), and  in part by the Center for Integrated Nanotechnologies --- a U.S. DOE Office of Basic Energy Sciences user facility (J.X.Z.).

\end{document}